\documentclass[5p]{elsarticle}
\makeatletter
\def\ps@pprintTitle{%
 \let\@oddhead\@empty
 \let\@evenhead\@empty
 \def\@oddfoot{}%
 \let\@evenfoot\@oddfoot}
\makeatother
\usepackage{lineno,hyperref}
\usepackage{amsmath} 
\usepackage[font={scriptsize}]{subcaption} 
\modulolinenumbers[1]
\graphicspath{{./figures/}} 
\journal{Nuclear Instruments and Methods in Physics Research A}

\begin{document}

\begin{frontmatter}

\title{Application of Bayes' theorem for pulse shape discrimination}

\author[UM_address]{Mateusz Monterial\corref{mycorrespondingauthor}}
\cortext[mycorrespondingauthor]{Corresponding author}
\ead{mateuszm@umich.edu}

\author[Sandia_address]{Peter Marleau}

\author[UM_address]{Shaun Clarke}

\author[UM_address]{Sara Pozzi}

\address[UM_address]{Department of Nuclear Engineering and Radiological Sciences, University of Michigan, Ann Arbor, MI 48104, USA}
\address[Sandia_address]{Radiation and Nuclear Detection Systems Division, Sandia National Laboratories, Livermore, CA 94551, USA}

\begin{abstract}
A Bayesian approach is proposed for pulse shape discrimination of photons and neutrons in liquid organic scinitillators. Instead of drawing a decision boundary, each pulse is assigned a photon or neutron confidence probability. This allows for photon and neutron classification on an event-by-event basis. The sum of those confidence probabilities is used to estimate the number of photon and neutron instances in the data. An iterative scheme, similar to an expectation-maximization algorithm for Gaussian mixtures, is used to infer the ratio of photons-to-neutrons in each measurement. Therefore, the probability space adapts to data with varying photon-to-neutron ratios. A time-correlated measurement of Am-Be and separate measurements of $^{137}$Cs, $^{60}$Co and $^{232}$Th photon sources were used to construct libraries of neutrons and photons. These libraries were then used to produce synthetic data sets with varying ratios of photons-to-neutrons. Probability weighted method that we implemented was found to maintain neutron acceptance rate of up to 90\% up to photon-to-neutron ratio of 2000, and performed 9\% better than decision boundary approach. Furthermore, the iterative approach appropriately changed the probability space with an increasing number of photons which kept the neutron population estimate from unrealistically increasing. 

\end{abstract}


\end{frontmatter}


\section{Introduction}

\subsection{Overview of the PSD problem}

In organic scintillators the fraction of light emitted during delayed fluorescence depends on the exciting particle, therefore it is possible to differentiate between neutron and photon interactions through pulse shape discrimination (PSD) \cite{Kno2000}. Although this property has been known for decades, recent advancements in pulse digitization \citep{Kornilov2003} and the demand for an alternative to the He-3 neutron detectors for security applications \cite{Shea2011} have invigorated the interest in PSD performance in organic scintillators. Reliable and robust PSD methods are necessary with organic scintillators to match the gamma rejection capabilities of He-3 detectors. 

The charge integration PSD method, in both analog and digital applications, relies on the ratio of pulse tail to total integrals \cite{Adams1978}. This PSD parameter is widely used in gauging PSD performance in organic scintillators \cite{Pawelczak2013, Liao2014, Pozzi2013}.  There are other methods for quantifying PSD parameters \cite{Gamage2011}, all of which provide a way of clustering neutrons and photons in a particular two dimensional space. Typically one dimension of this space is the PSD parameter, and the other is some metric of energy deposition such as pulse height or pulse integral. A well chosen PSD parameter ensures adequate separation between photon and neutron populations. The width of the photon and neutron distributions is inversely proportional to deposited energy, which makes classification difficult at lower energies. In this paper we apply a low threshold of 25 keVee ("kilo-electron Volt electron equivalent"), equivalent to estimated 245 keV neutron deposited energy. 

\subsection{New classification method} \label{sec:ncm}

In this work we present a new classification methodology which departs from typical approaches of drawing the optimal decision boundary between photon and neutron distributions \cite{Kaplan2013,Polack2013}. In our methodology, instead of segregating pulses into distinct groups, each pulse is assigned a neutron and photon weight based on posterior probability calculated from Bayes' theorem using the photon-to-neutron ratio as the prior probability as will be discussed in Section \ref{sec:itpnr}. The posterior probabilities are then used to estimate a new prior, and the process repeats until a convergence criteria is met. This iterative scheme successfully adapts the probability space to different proportions of neutrons and photons in the measured data. Furthermore, the sum of posterior probabilities is shown to be a better estimator of total instances of photons and neutrons than an optimal decision boundary that minimizes misclassifications. For our two class problem, the decision boundary that minimizes misclassification is set at a line where posterior probability equals 50\%. Events with a PSD parameter above this boundary are classified as neutrons and below as photons. 

Performance of the PSD method in this paper was quantified from a neutron detection point of view. Any effective PSD method has to simultaneously maintain high neutron efficiency and a high photon rejection rate \cite{Kauzes2009}. Photon data sets were taken from measurements of pure photon sources, and neutrons from a time correlated measurement of an Am-Be source. Photon and neutron pulse data were combined to form synthetic data sets with varying photon-to-neutron ratios.  Neutron efficiency was calculated as a function of this ratio by summing posterior probabilities, we define this as the probability weighted method. This was then compared to neutron efficiency obtained by discrimination using a decision boundary. 

\section{Experiment}

Multiple measurements were performed with the same experimental setup to benchmark the new Bayesian PSD method. An Am-Be source was measured to provide a data set of time tagged neutrons. Three photon sources, $^{137}$Cs, $^{60}$Co and $^{232}$Th were measured separately to provide the corresponding set of pure photon pulses. The photon and neutron data sets were combined to produce mixed data sets with desired photon-to-neutron ratios. 

This approach was favored over combined measurement of neutron and photon radiation sources for two reasons. First, direct measurement of high photon-to-neutron ratio would necessitate high count rates, which introduces complications from incomplete double pulse rejection. Double pulses are readily characterized as neutron pulses, and their inclusion perturbs the metrics used to quantify PSD performance. Secondly, synthetically adding in photon pulses into the mixed data sets gives us greater control in the choice of a photon-to-neutron ratio. Finally, uncertainties associated with the detector-source geometry and source strength, which would be included if mixed radiation fields were measured directly, are precluded from the estimation of photon-to-neutron ratio when data sets are mixed after measurements.

\subsection{Set-up and detector settings}

Two 2$\times$ 2" EJ-309 organic liquid scintillators coupled to Hamamatsu H1949-50 photomultiplier tubes (PMT) were used for the measurements. Full waveforms were digitized with a CAEN DT5720, 12-bit, 250 MHz digitizer which were saved on a computer through a USB cable. For the Am-Be measurement, one detector was positioned approximately 49.5 cm from the source and was used to benchmark the PSD method; a second detector was positioned approximately 0.5 cm from the source and was used to time tag correlated fast neutrons and high energy gammas originating in the ($\alpha$,n) reaction \cite{Vega2001}. The biases applied on the first and second PMTs were -2050 V and -1800 V, respectively. Despite the disparity in the applied voltage, the effective gains were similar because of the use of a 20 dB and 14 db attenuators with the first and second detectors, respectively. 

\begin{figure}[h]
	\centering
	\includegraphics[width=8.6cm]{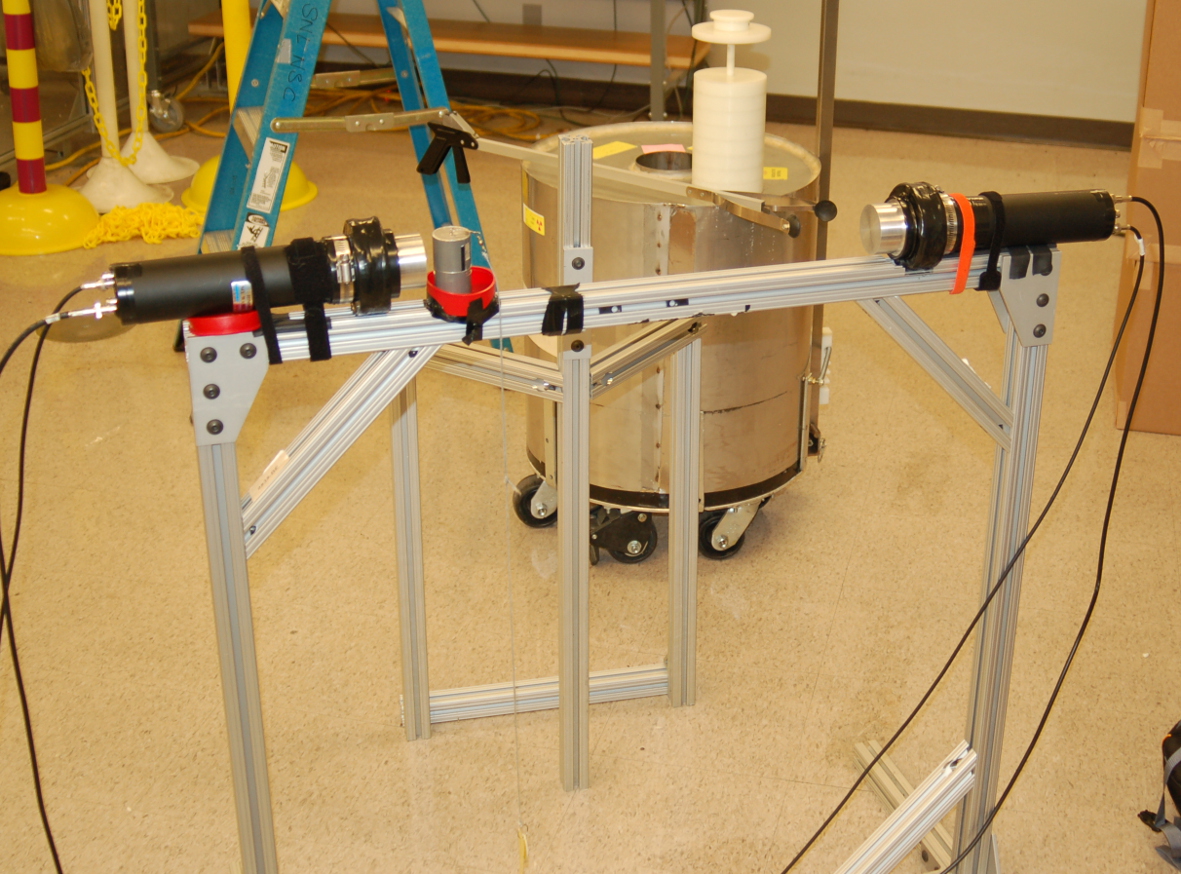}
    \caption{Setup of the two detectors with Am-Be source in position.}
    \label{fig:setup}
\end{figure}

The two detectors were configured in a face-to-face orientation 50 cm apart on an aluminum frame 100 cm from the floor. The 73.7 mCi Am-Be source was placed 0.5 cm away from the second detector for a 108 hour long measurement. The three photon sources measured were 45.7 $\mu$Ci $^{137}$Cs, 54.5 $\mu$Ci $^{60}$Co and 13.5 $\mu$Ci $^{232}$Th. These photon sources were measured 20, 25, and 15 cm away from the first detector, respectively. The photon sources were measured at different distances to keep the count rate constant around 2000 Hz. A picture of the setup with the Am-Be source in position is shown in Figure \ref{fig:setup}. 
 
The position of the Compton edge was estimated at 80\% of the peak height, through linear interpolation between edge peak and baseline. Pulse waveforms were integrated with a 240 ns integration window. The effective measurement threshold was ~18 keVee, but 25 keVee threshold was used in post processing. This threshold corresponds to neutron energy deposited on hydrogen of 245 keV, as calculated from EJ-309 integral light output coefficients \cite{Lawrence2014}.

\section{Method}

Applying Bayes' theorem requires a conditional probability, or likelihood, and a prior probability. In this work, we define the former as the value of an energy dependent Gaussian fit for a given tail-to-total ratio and the latter as the energy dependent photon-to-neutron ratio. The discussion on estimating both parameters proceeds in the following two sections. Initially, the prior probability is not known and thus it must be inferred from the data by an iterative procedure. Therefore, the Bayesian probability is adaptive to a particular collection of data. 

\subsection{Fitting detector specific parameters}

The detector specific parameters were determined from an energy-dependent double Gaussian fit to the tail to total ratio vs. total integral histogram. This step would preferably be accomplished on a data-set with nearly equal photon and neutron populations. The fit provides energy dependant means and standard deviations for both photons and neutrons. These parameters are only system dependent, and like gain calibration procedures only need to be performed once for a particular experiment. 

An open source unconstrained non-linear optimization algorithm was used for fitting the Gaussian parameters \cite{PeakFit}. Each group was fit in descending order of light output, with the previously calculated coefficients used as the next initial guess. This ensured continuity of the coefficients across all groups, and facilitated the precarious fitting at lower light outputs where distributions overlap the most. The Gaussian distributions were normalized and took on the familiar form 

\begin{align} \label{eq:gauss}
f(s) = \frac{1}{\sigma \sqrt{2\pi}} \exp\left(-\frac{(s-\mu)^2}{2\sigma^2}\right)
\end{align} 
where $s$ is the tail-to-total ratio.

\begin{figure}[h]
	\centering
	\includegraphics[width=8.6cm]{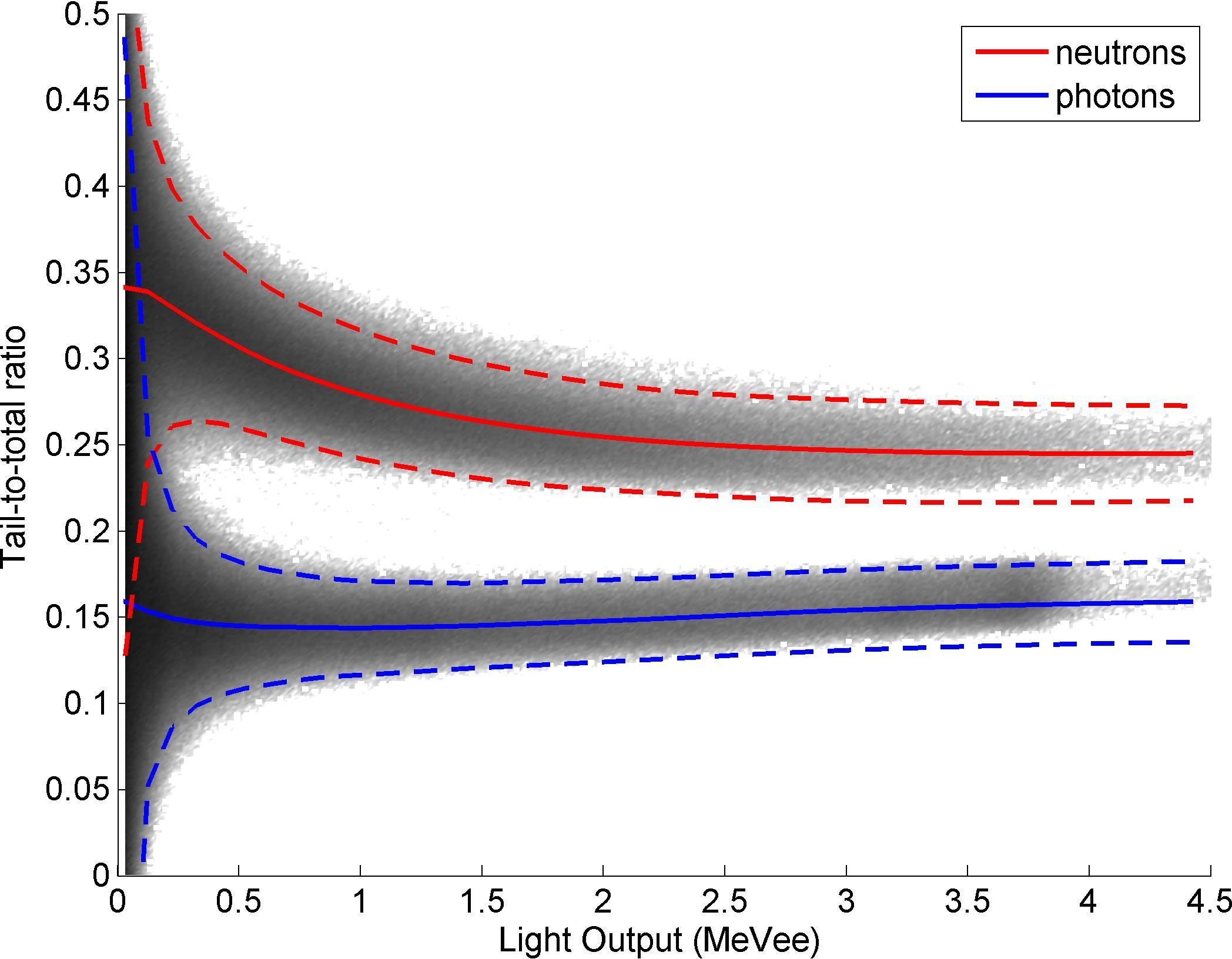}
    \caption{The mean (solid lines) and three standard deviation (dashed lines) fits as applied to the Am-Be data set.}
        \label{fig:param_fit}
\end{figure}

In order to calculate the Bayesian probability for each pulse individually, the Gaussian coefficients were fit across light output range of interest. Smoothing splines were used to approximate the mean and standard deviation coefficients as a function of light output. The resulting R-squared values were greater than 0.99 for all fitted parameters. The result of the fitting procedure is shown in Figure \ref{fig:param_fit}, as applied to a subset of the Am-Be data. Conditional probabilities used for PSD analysis in this paper were taken from these fits. 

By fixing the mean and standard deviation parameters from a calibration data set we assume they only depend on the detector system, and therefore do not need to be refitted for other measurements. A comparison of mean and standard deviation fits for the photon sources measured,  $^{137}$Cs, $^{60}$Co and $^{232}$Th, has shown this to be the case. In all three cases the standard deviations were nearly identical across the entire light output range. The mean varied near regions of high count rate (i.e., Compton edge), although at most only by 10\% of the standard deviation. As a result it is possible to use a single calibration data set --- preferably one with nearly equal number of photons and neutrons --- to produce the necessary fits. 

\subsection{Inferring the photon-to-neutron ratio} \label{sec:itpnr}

The Gaussian fits provide the conditional probabilities, but we still require a prior in order to calculate a photon or neutron posterior probability. In our case this is the energy dependant ratio of photon-to-neutrons, which will change depending on the incident radiation (i.e. the type of radiation source measured). Therefore we introduce an iterative procedure that updates the photon-to-neutron ratio by recalculating posterior probabilities until a convergence criteria in the photon and neutron populations is met. 

The following formulation of the Bayes' theorem was used for calculating posterior probabilities for either photons

\begin{align} \label{eq:p_gr}
P(\gamma|s) = \frac{f_\gamma(s)R_{\gamma/n}}{f_\gamma(s)R_{\gamma/n}+f_n(s)}
\end{align} 
or neutrons
\begin{align} \label{eq:p_nr}
P(n|s) = \frac{f_n(s)}{f_\gamma(s)R_{\gamma/n}+f_n(s)}
\end{align}
where $f(s)$ are the Gaussian fits from Eq. \ref{eq:gauss}, the $\gamma$ or $n$ index indicates the photon and neutron distribution and $s$ is the PSD parameter, in this case the tail-to-total ratio. $R_{\gamma/n}$ is the ratio of the estimated number of counts in the photon and neutron distributions within a light output group. The number of instances of photons and neutrons is estimated by summing the posterior probabilities of each class
\begin{align}
N_\gamma &= \sum\limits_{s \in E_i}{P(\gamma|s)} \label{eq:Ng} \\
N_n &= \sum\limits_{s \in E_i}{P(n|s)} \label{eq:Nn}
\end{align}
for a particular light output group $E_i$.

For the first iteration $R_{\gamma/n}$ is assumed to be unity, then the results from Eqs. \ref{eq:p_gr} and \ref{eq:p_nr} are used to estimate its value for the subsequent iteration:
\begin{align} \label{eq:R}
R_{\gamma/n} = \frac{N_\gamma}{N_n}.
\end{align}

The iterative approach used is an example of an expectation-maximization algorithm for Gaussian mixtures \cite{Uchida2014} with fixed means and standard deviations. Iterations terminate when the convergence criteria is satisfied. Convergence criteria was defined as 1\% difference in total $R_{\gamma/n}$ between two consecutive iterations. This iterative scheme is robust, and converges to the same solution given a wide range of initial $R_{\gamma/n}$. 

\subsection{Variance estimation}

Since the sum of probabilities, not counts, is used to estimate the size of photon and neutron populations, we added a term in the variance calculation to account for the uncertainty of the posterior probability itself.  In this work the variance on the total number of estimated instances $N$ was calculated by

\begin{align} \label{eq:var}
Var(N) = \sum{P_i} + \sum{P_i(1-P_i)}
\end{align} 
where $P_i$ are either photon or neutron posterior probabilities for each pulse. $N$ represents the number of photon or neutron instances as calculated by Eqs. \ref{eq:Ng} and \ref{eq:Nn}. The first term in Eq. \ref{eq:var} sums to $N$ and represents the variance on the total number of counts. The second term is the variance on each individual probability, which follows a binomial distribution because there are only two classes, photons and neutrons. In the limit that all the $P_i$ are large the variance estimation reduces to $N$. On the other hand, if $P_i$ are small than the variance estimation is effectively doubled to $2N$. 

\section{Results}

\subsection{Time-tagged neutrons}

\begin{figure}[h!]
	\centering
	\includegraphics[width=8.6cm]{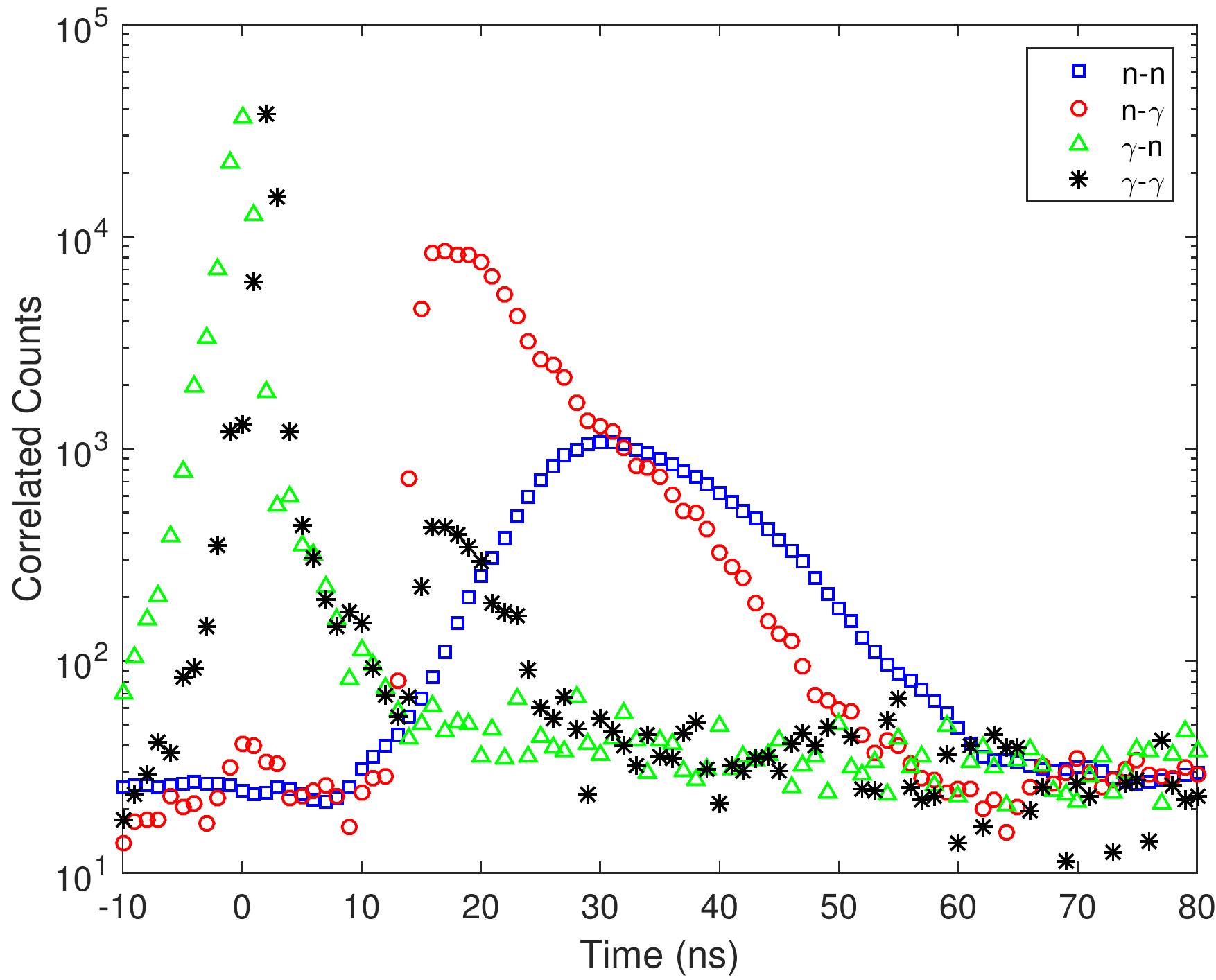}
    \caption{Timing distribution of correlated counts between two liquid scintillator detector using and Am-Be source. The minimum correlation probability threshold was 99\% to minimize appearance of misclassified correlations.}
        \label{fig:cross_corr}
\end{figure}

A set of ``pure neutron" pulse data was needed to benchmark the Bayesian PSD method, and it was obtained from time correlated measurement of the Am-Be source. Figure \ref{fig:cross_corr} shows the resulting correlated counts for all particle pair combinations weighed by correlation probabilities. Eqs. \ref{eq:p_gr} and \ref{eq:p_nr} were used to calculate the Bayesian probabilities for each detector.

The presence of photon-photon correlations at times greater than 12 ns is due to the production of photons induced by high energy neutrons. Only one correlated neutron is produced from the ($\alpha$,n) reaction, however the presence of neutron-neutron correlations indicates significant contributions from both scatter and (n,2n) reactions in Beryllium.
	
The neutron data set was taken from the pulses with time correlations between 15 and 50 ns, which corresponds to neutron energies of ~5.6$\pm$0.4 MeV and 460$\pm$19 keV given detector separation of 50 cm. We approximated 96.5\% neutron purity, for a total of 105,000 total neutron counts, in the time-tagged neutron data set by applying the PSD approach presented in this paper. 

\subsection{Neutron acceptance}
   
\begin{figure}[h]
	\centering
	\includegraphics[width=8.6cm]{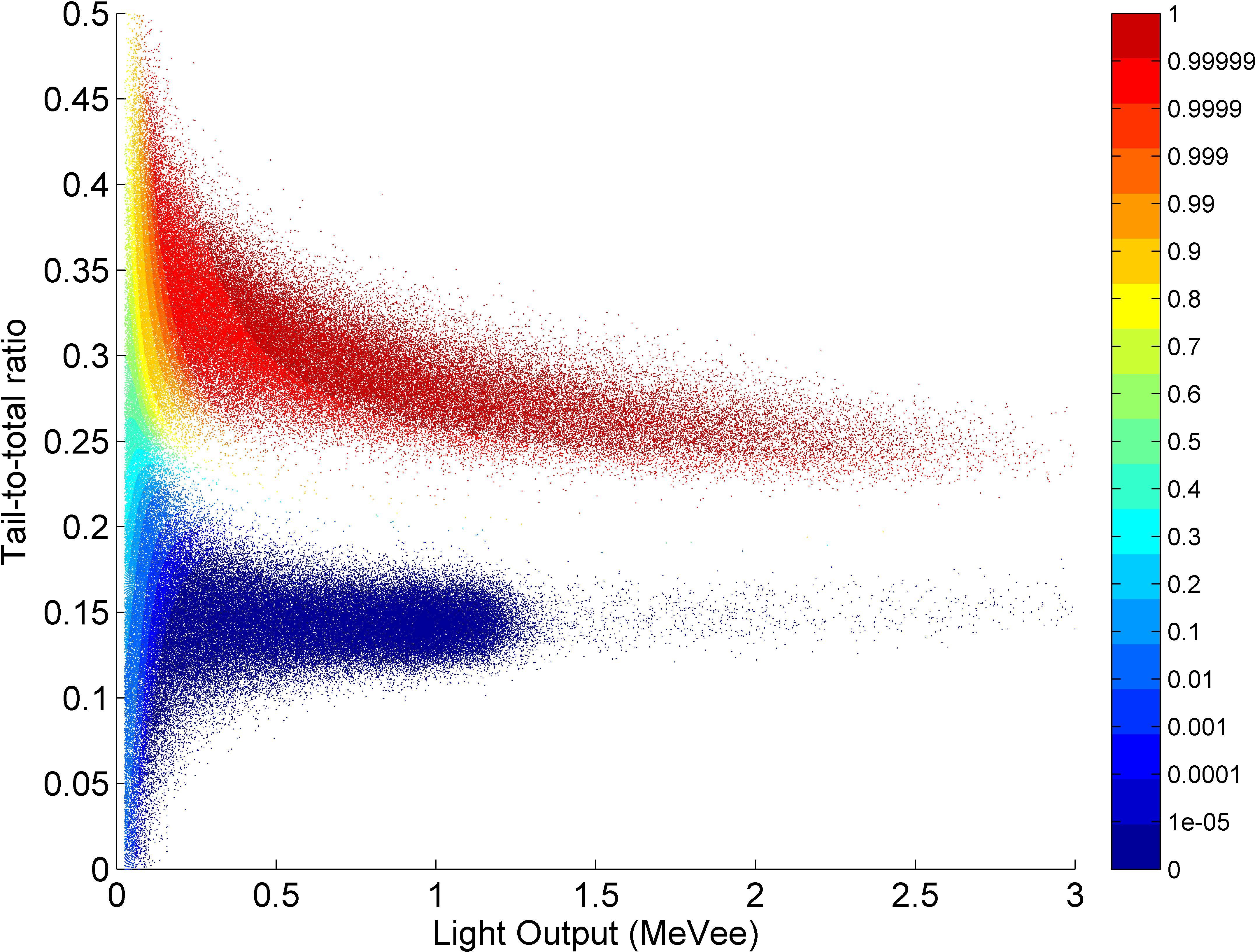}
    \caption{The distribution of neutron posterior probabilities for 1:1 mixture of $^{60}$Co and time tagged neutron data (color available online).}
        \label{fig:neutronicity}
\end{figure}

The neutron acceptance rate is defined relative to the counts found within the timing window in the Am-Be measurement before mixing in photons from the gamma source data sets. Photon pulses from  $^{137}$Cs, $^{60}$Co and $^{232}$Th measurements were mixed into the neutron data with estimated total photon-to-neutron proportions ranging from 1 to 2000. An example of the neutron probability map from a 1:1 mixture with $^{60}$Co data is shown in Figure \ref{fig:neutronicity}. Our iterative approach ensures that the neutron probability decreases as more photons are added to the mixed set. 

\begin{figure*}[h!]
        \centering
        \begin{subfigure}[h]{0.34\textwidth}
			\centering
			\includegraphics[width=\textwidth]{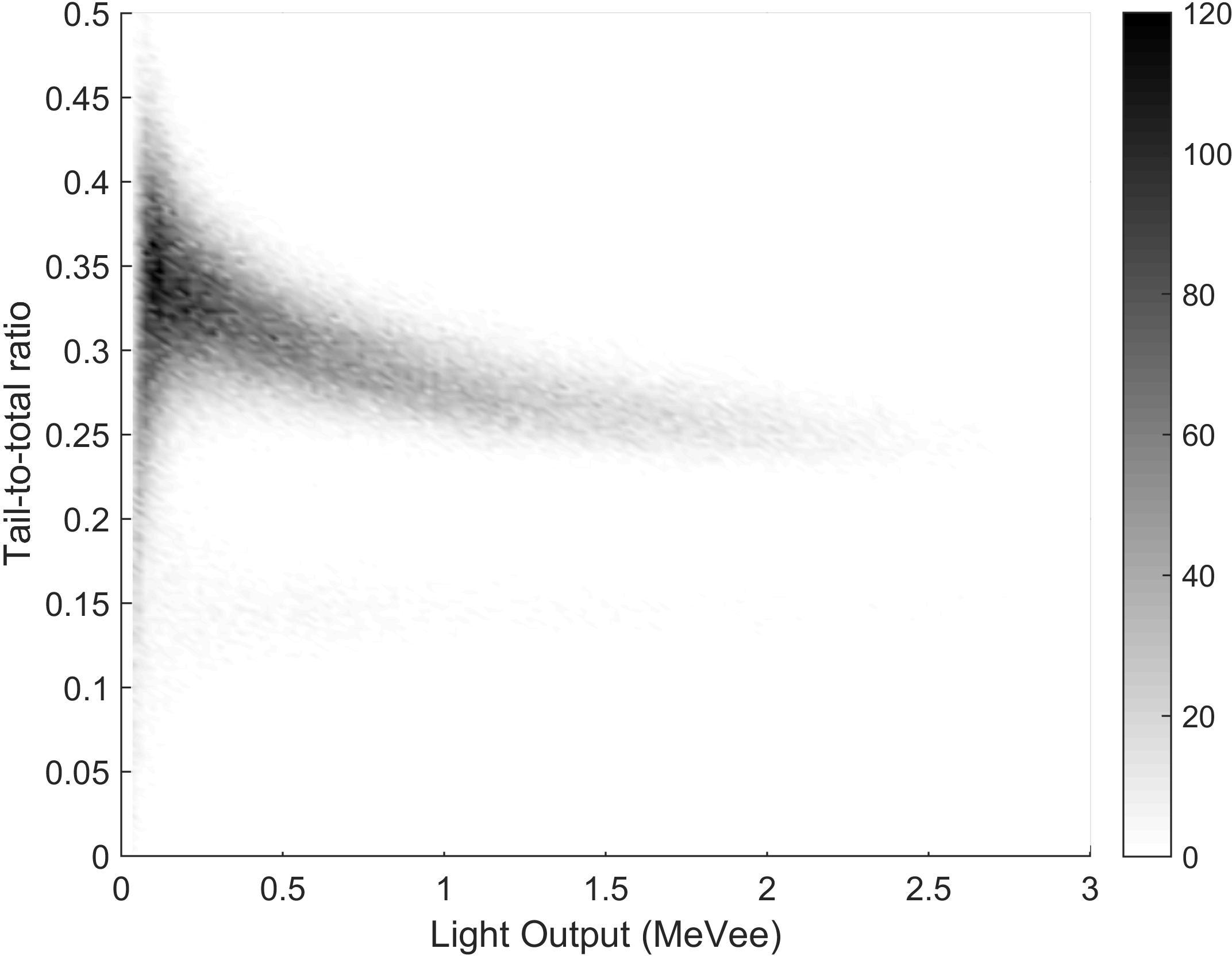}
    		\caption{Neutrons before}
        \end{subfigure}%
        ~ 
        \begin{subfigure}[h]{0.34\textwidth}
			\centering
			\includegraphics[width=\textwidth]{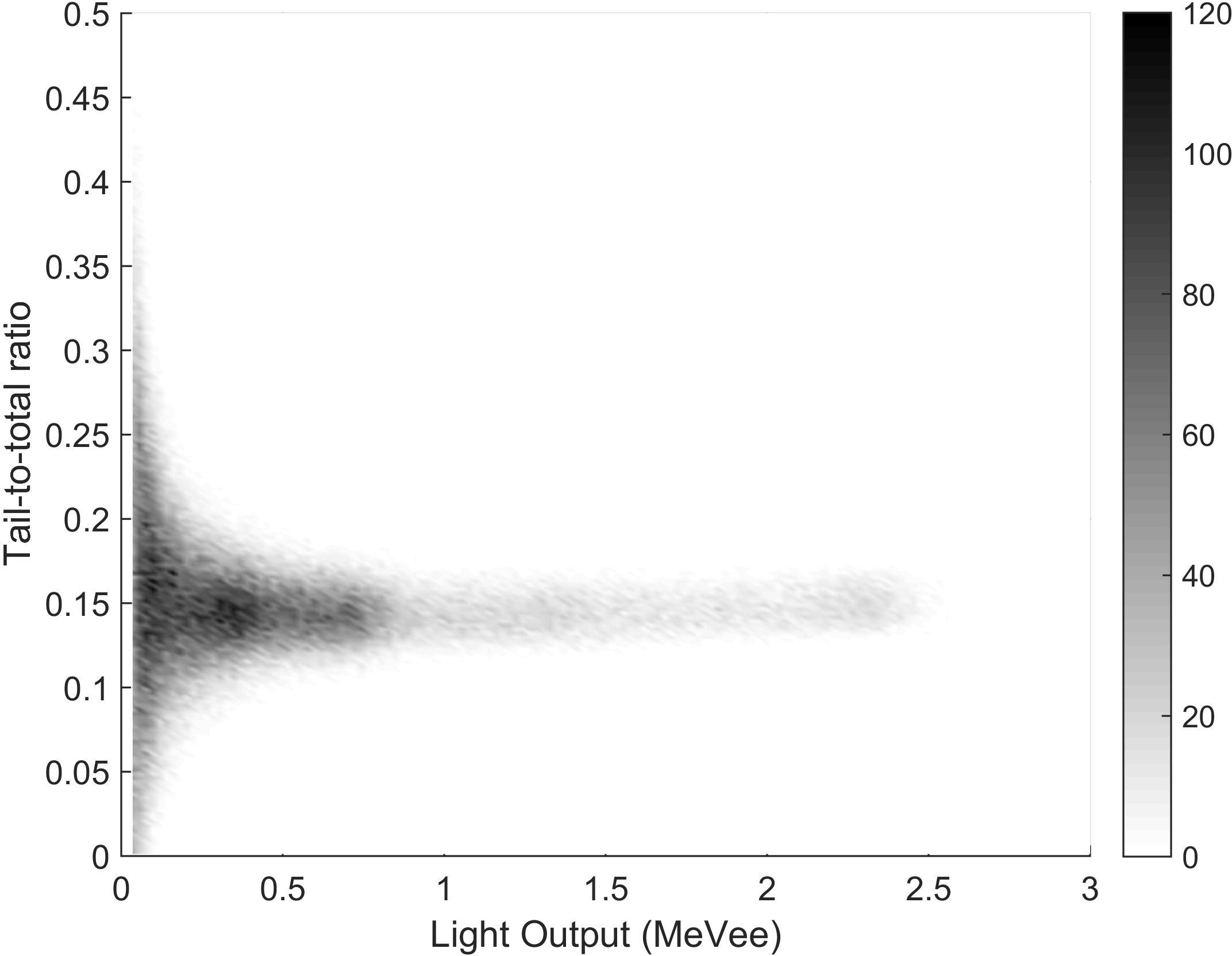}
    		\caption{Photons before}
        \end{subfigure}%
        
                \centering
        \begin{subfigure}[h]{0.34\textwidth}
			\centering
			\includegraphics[width=\textwidth]{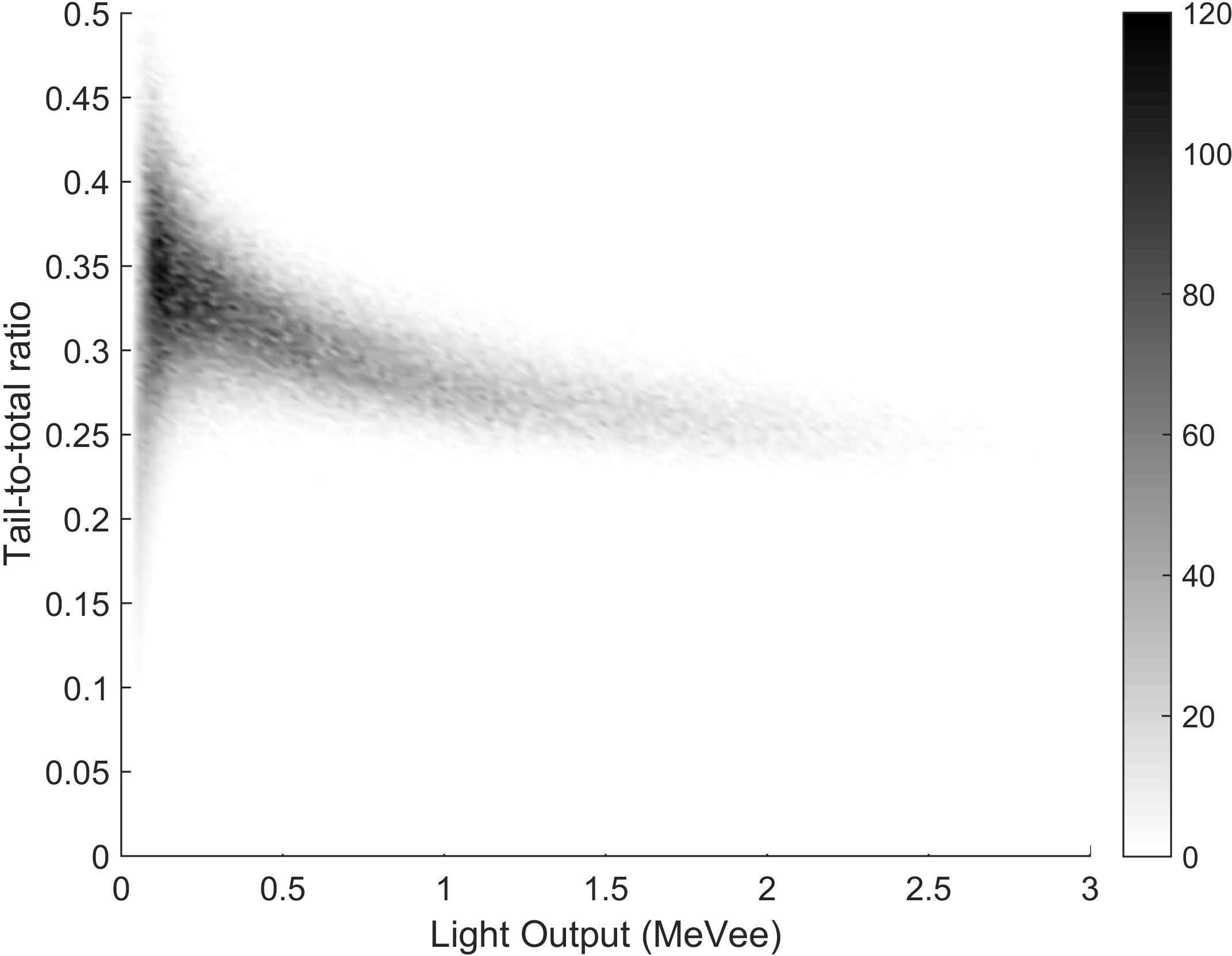}
    		\caption{Neutrons after}
        \end{subfigure}%
        ~ 
        \begin{subfigure}[h]{0.34\textwidth}
			\centering
			\includegraphics[width=\textwidth]{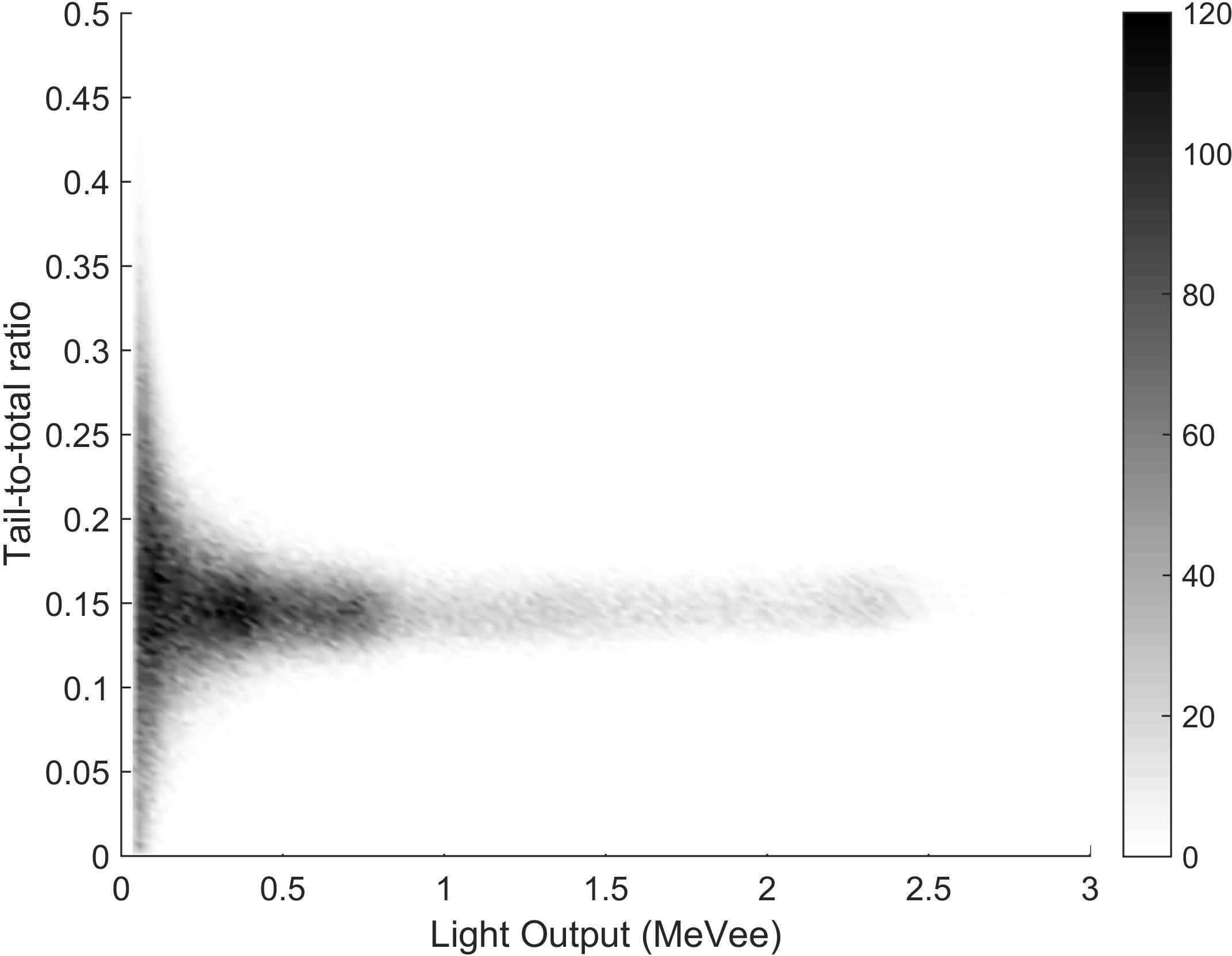}
    		\caption{Photons after}
        \end{subfigure}%
        
        \caption{The distribution of time-tagged neutrons and $^{232}$Th photons before and after mixing in a 1:1 ratio.}
        \label{fig:Th_mix_1}
\end{figure*}

\begin{figure*}[h!]
        \centering
        \begin{subfigure}[h]{0.34\textwidth}
			\centering
			\includegraphics[width=\textwidth]{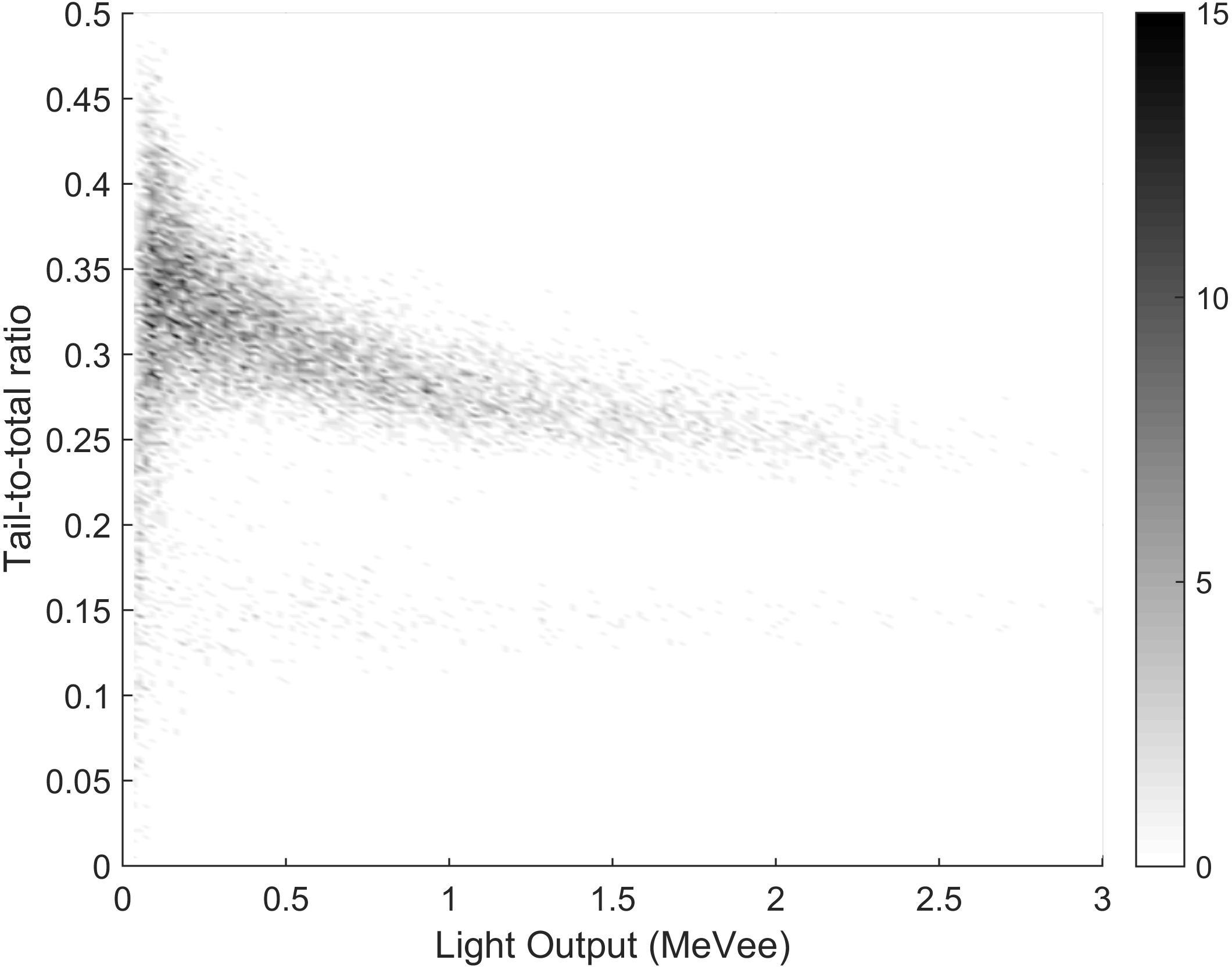}
    		\caption{Neutrons before}
        \end{subfigure}%
        ~ 
        \begin{subfigure}[h]{0.34\textwidth}
			\centering
			\includegraphics[width=\textwidth]{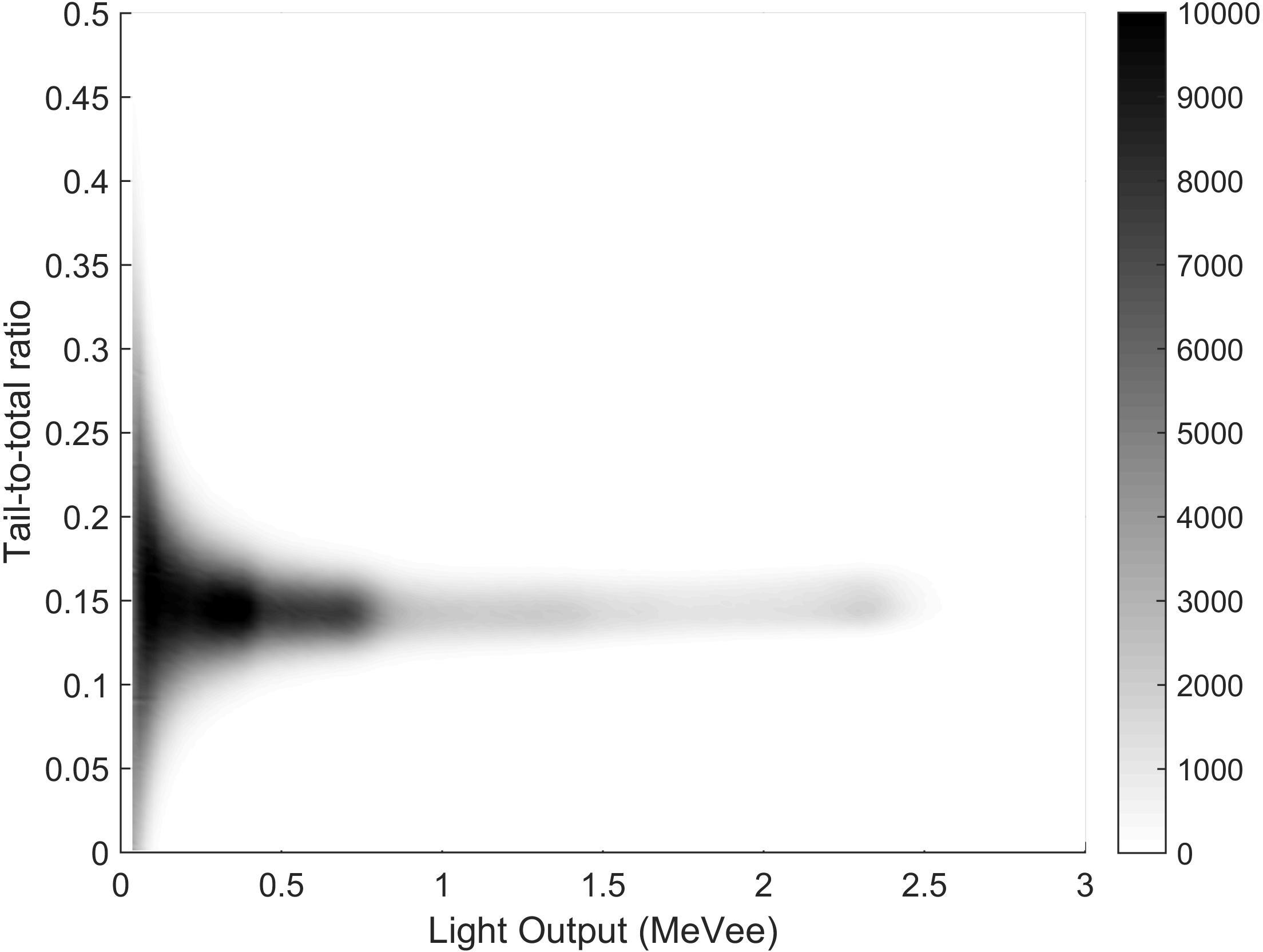}
    		\caption{Photons before}
        \end{subfigure}%
        
                \centering
        \begin{subfigure}[h]{0.34\textwidth}
			\centering
			\includegraphics[width=\textwidth]{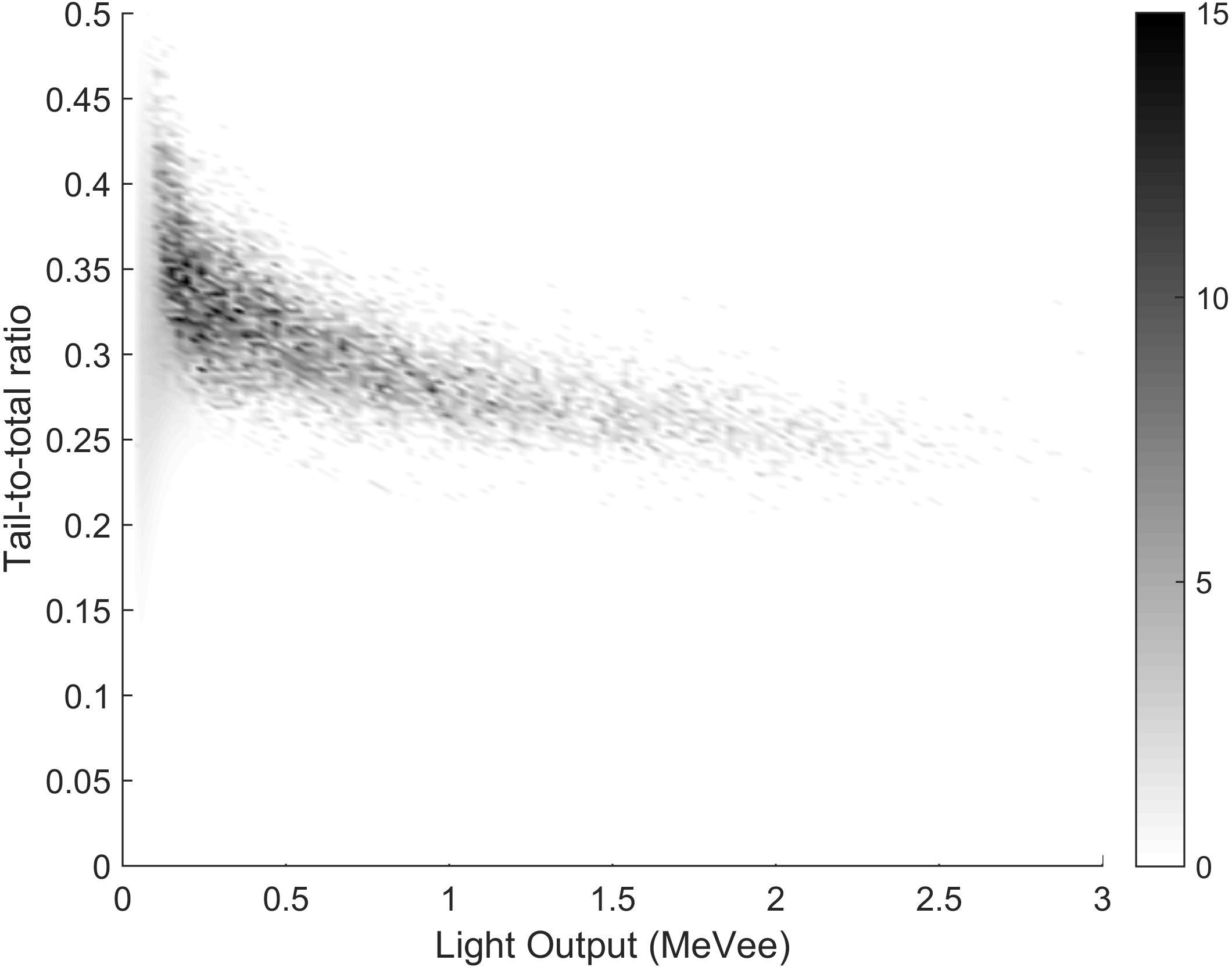}
    		\caption{Neutrons after}
        \end{subfigure}%
        ~ 
        \begin{subfigure}[h]{0.34\textwidth}
			\centering
			\includegraphics[width=\textwidth]{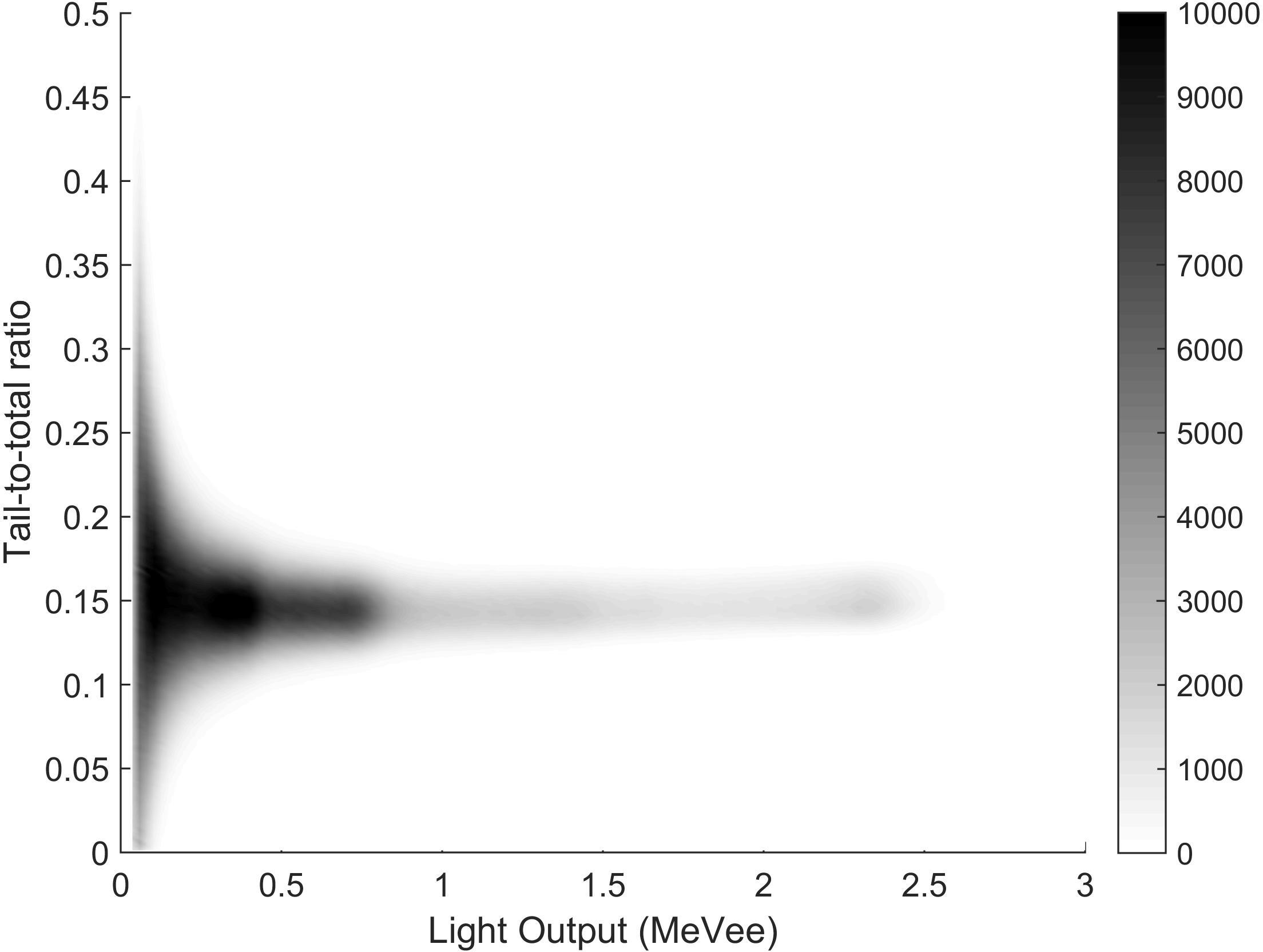}
    		\caption{Photons after}
        \end{subfigure}%
        
        \caption{The distribution of time-tagged neutrons and $^{232}$Th photons before and after mixing in a 1000:1 ratio.}
        \label{fig:Th_mix_1000}
\end{figure*}

\begin{figure}[h]
	\centering
	\includegraphics[width=8.6cm]{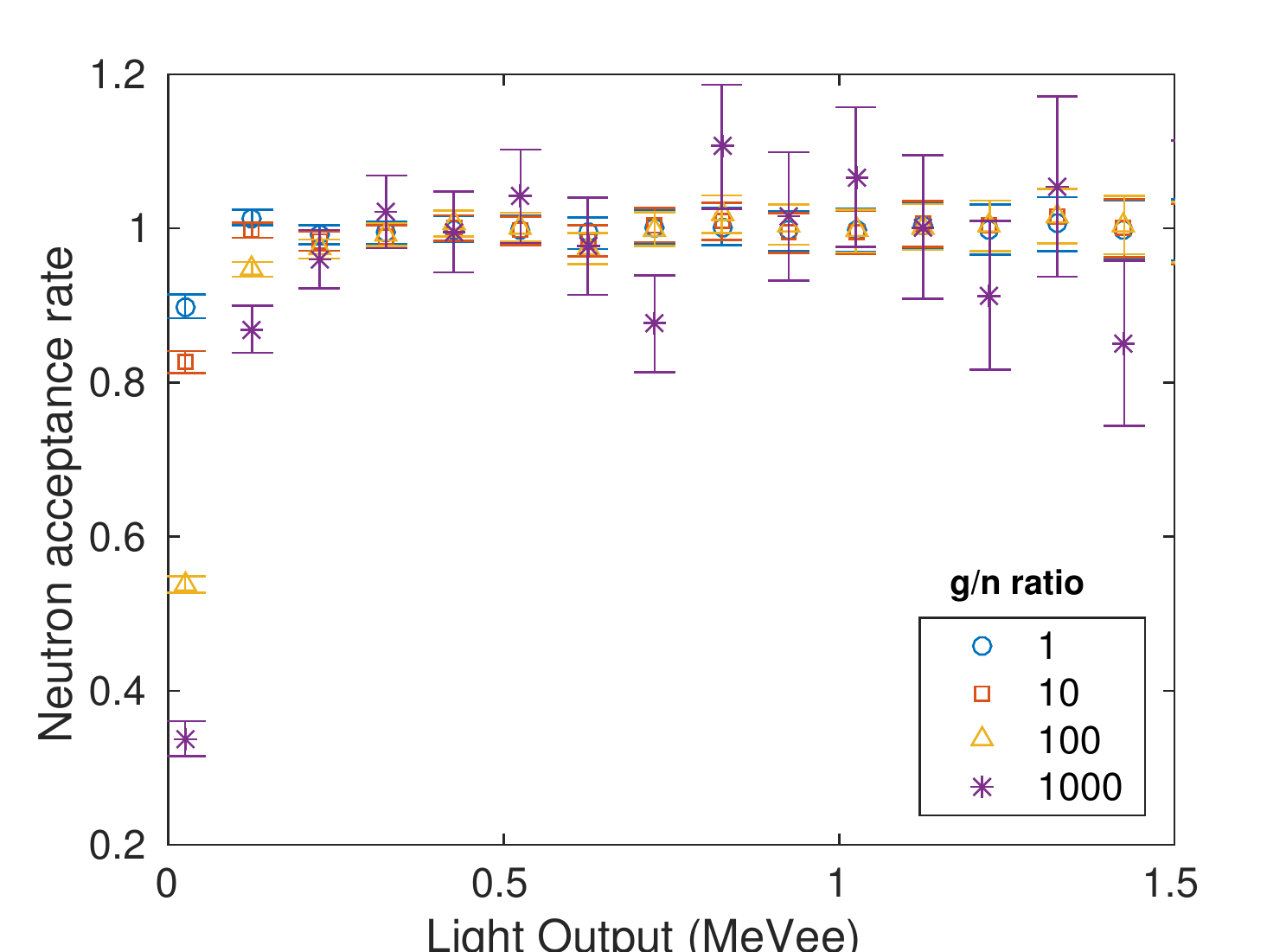}
    \caption{Neutron acceptance rate for the  $^{232}$Th data set as a function of light output for different photon-to-neutron ratios.}
        \label{fig:n_eff_erg}
\end{figure}

Figures \ref{fig:Th_mix_1} and \ref{fig:Th_mix_1000} show the photon and neutron distribution before being mixed and after separation for 1:1 and 1000:1 photon-to-neutron ratios, respectively. In the 1:1 mixture the neutron distribution shape and magnitude is persevered after mixing, however the number of photons has increased because of the photons present in the time-tagged neutron data set. In the 1000:1 mixture the contribution of those photons is relatively much smaller, and as a result the photon distributions before and after mixing appear nearly identical. Due to the higher number of photons present in the 1000:1 mixture, the neutron probabilities were reduced causing a drop in efficiency, particularity at low energies. The drop in neutron acceptance rate at low energies is demonstrated in Figure \ref{fig:n_eff_erg}, where neutron acceptance rate is presented for each light output group. Only 10\% of the available neutrons, approximately 10,500 out of the set of 105,000, were used in the 1000:1 mixture because of the limited number ($\sim$15 million) of photon events available from the gamma source measurements. 

\begin{figure}[h]
        \centering
        \begin{subfigure}[h]{0.45\textwidth}
			\centering
			\includegraphics[width=6.5cm]{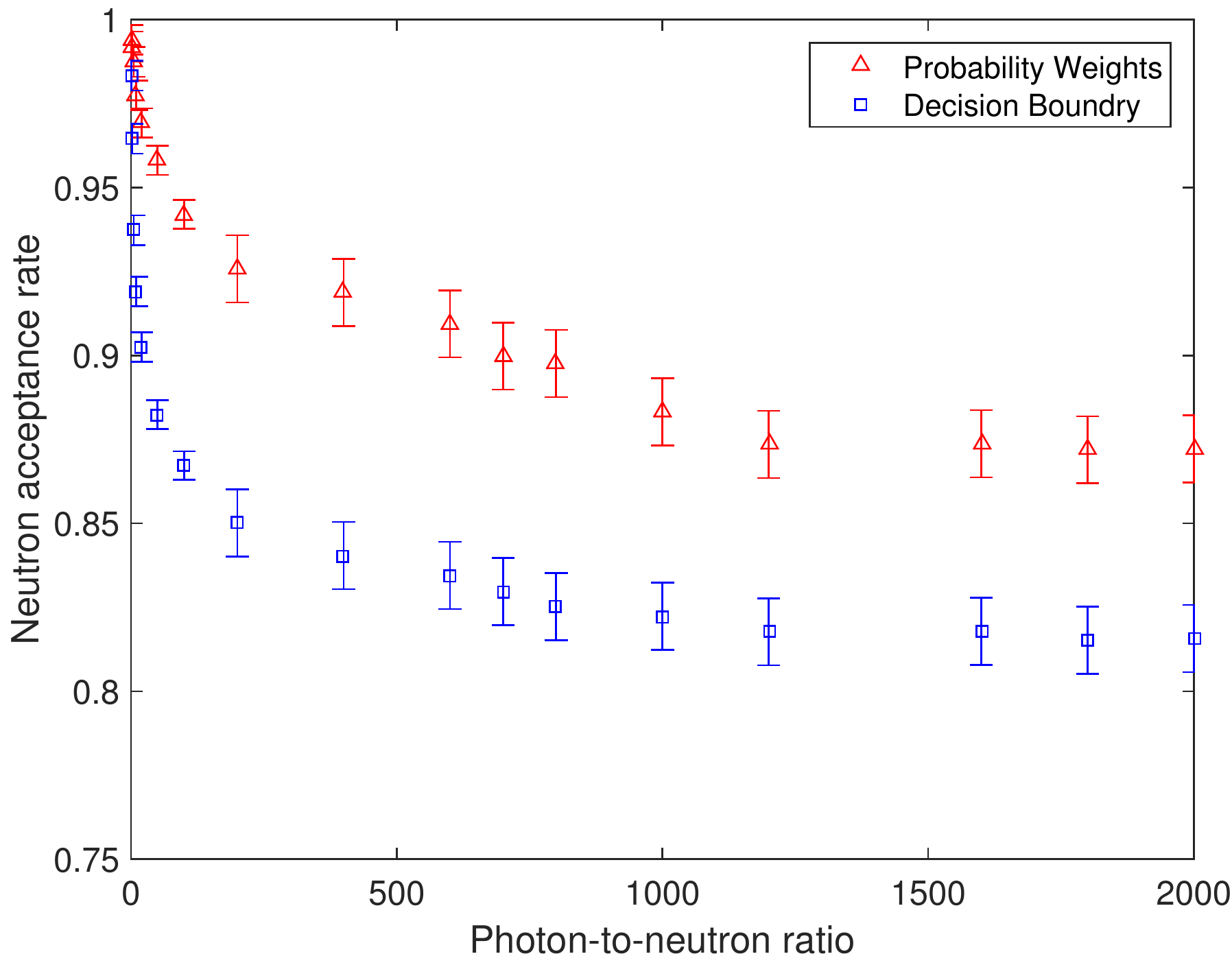}
    		\caption{$^{137}$Cs}
        \end{subfigure}%
        
        ~ 
        \begin{subfigure}[h]{0.45\textwidth}
			\centering
			\includegraphics[width=6.5cm]{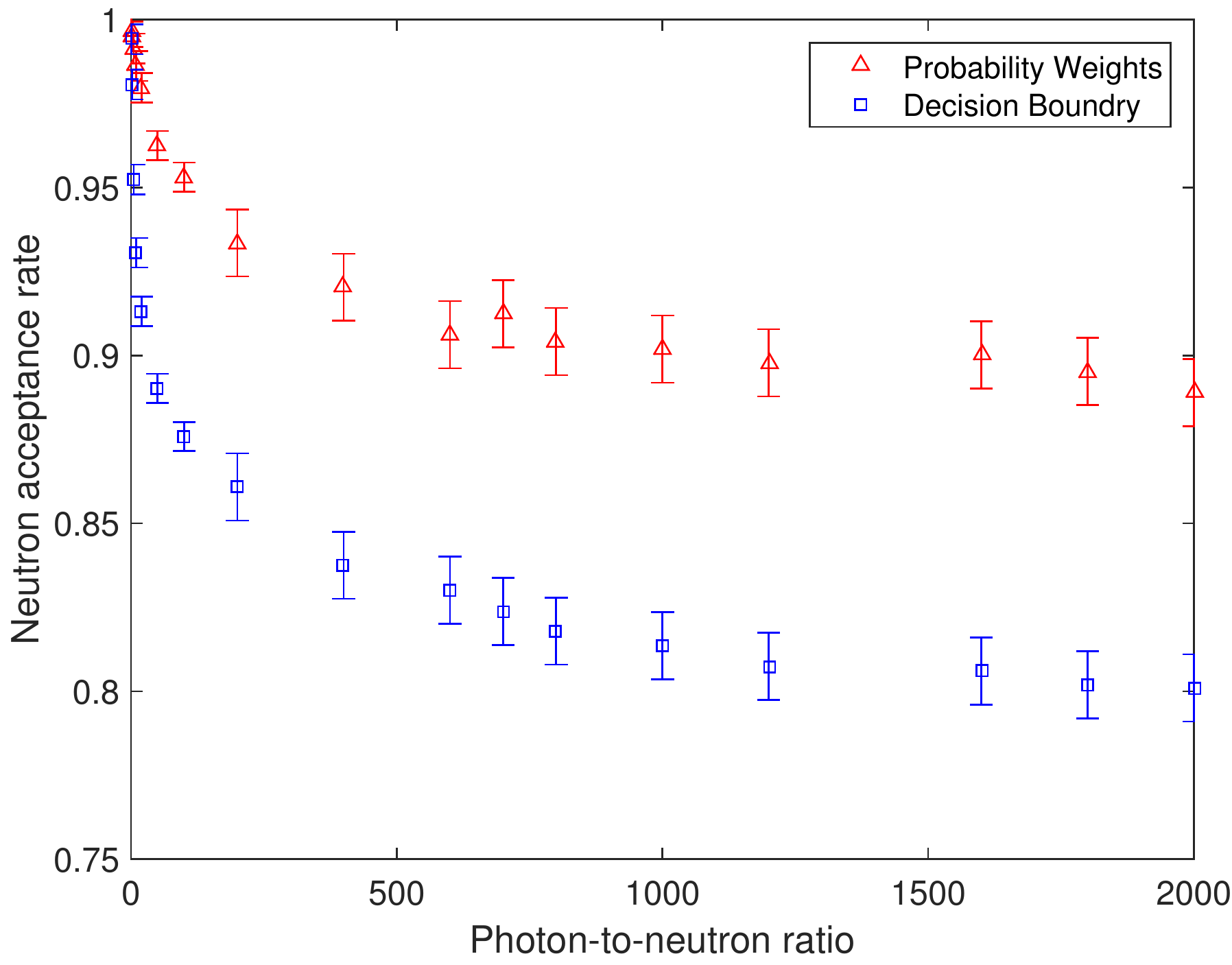}
    		\caption{$^{60}$Co}
        \end{subfigure}%
      
      ~  
        \begin{subfigure}[h]{0.45\textwidth}
			\centering
			\includegraphics[width=6.5cm]{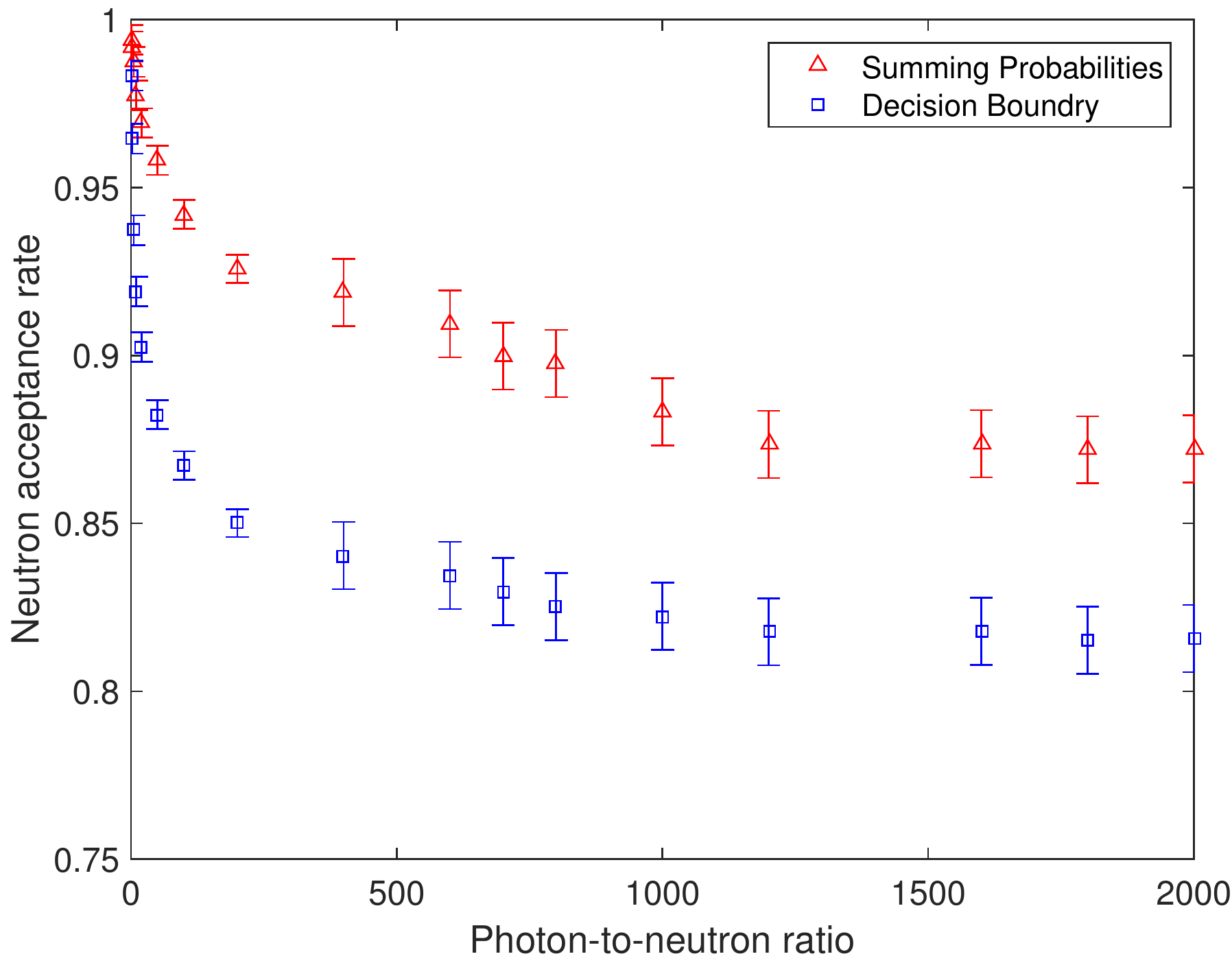}
    		\caption{$^{232}$Th}
        \end{subfigure}%
                
        \caption{Neutron acceptance rate as a function of photon-to-neutron ratio for all three photon sources. Only 10\% of the available neutron had to be used to achieve ratios greater than 100, which results in larger error estimations.}
        \label{fig:n_acc}
\end{figure}

The neutron acceptance rate as a function of photon-to-neutron ratio is shown in Figure \ref{fig:n_acc}. The neutron acceptance rate shown is integrated over all light output groups, essentially the integrated result shown in Figure \ref{fig:n_eff_erg} but for all gamma sources. The neutron counts were weighted by their neutron posterior probabilities, from Eq. \ref{eq:p_nr}, which we call the probability weights technique. Since neutron posterior probabilities are calculated for every event we can sum them to estimate the total population of neutrons. This was compared to the integral of counts above the decision boundary defined in Section \ref{sec:ncm}, where every event is counted as either a neutron or a photon. For both methods the same iterative procedure, described in Section \ref{sec:itpnr}, was used to approximate the prior distribution $R_{\gamma/n}$. As expected, the summation of posterior probabilities maintains an acceptance rate up to 9\% higher than those obtained using decision boundary discrimination. 

\begin{figure}[h]
	\centering
	\includegraphics[width=8.6cm]{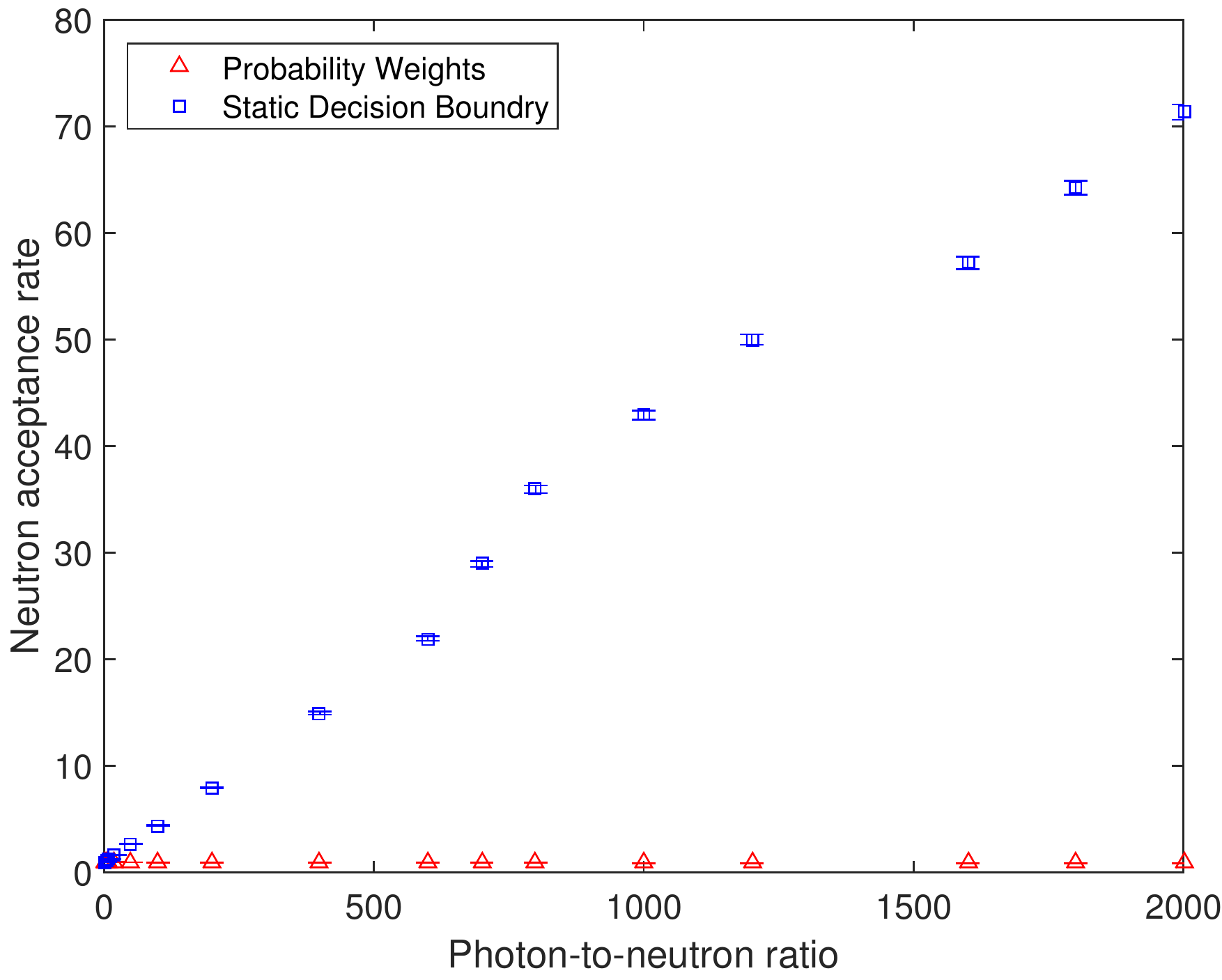}
    \caption{Neutron acceptance rate for the $^{137}$Cs data set with the static decision boundary chosen assuming photon-to-neutron ratio of 1.}
        \label{fig:n_acc_Cs_static}
\end{figure}

Regardless of the photon-to-neutron ratio, the integrated neutron acceptance rate was kept below 100\% because of the iterative approach that adjusted the photon-to-neutron ratio for each data set. At some light output values the acceptance rate did exceed 100\% but only within a few percent of the marginal error. Traditionally the decision boundary is set based on a calibration data set and is then used for subsequent measurements \cite{Kaplan2013}. This traditional approach is compared to our iterative technique in Figure \ref{fig:n_acc_Cs_static}. The static decision boundary was chosen assuming a photon-to-neutron ratio of 1. The problem with using a static boundary is that photons always have some probability of being above it in the neutron region. The neutron acceptance rate will increase linearly with the photon-to-neutron ratio as more photons are added into the mix. 

\clearpage

\section{Conclusions}
The applicability of Bayes' theorem to pulse shape discrimination was demonstrated with the use of a charge integration technique. An iterative approach that infers the ratio of photon-to-neutron light output distributions from data is demonstrated as a robust method for adjusting the probability space for measurements with different photon-to-neutron ratios. Neutron acceptance rate of 90\% was maintained with photon-to-neutron ratio of up to 2000. Overall the neutron acceptance rate was up to 9\% higher with the probability weighted technique compared with drawing a decision boundary. 

The Bayesian PSD technique can be used to calculate neutron and photon probabilities on an event by event basis. For correlation experiments with multiple detectors, correlation probabilities for each possible event type combination can be calculated. These correlation probabilities can be used to set new thresholds to clean-up misclassification rates to a desired level without raising the energy threshold on all events.    

\section*{Acknowledgments} 

The author would like to thank Kyle Polack, Marek Flaska, and Aaron Bevill for their feedback regarding the work presented in this paper.

This material is based upon work supported by the U.S. Department of Homeland Security under Grant Award Number, 2012-DN-130-NF0001. The views and conclusions contained in this document are those of the authors and should not be interpreted as representing the official policies, either expressed or implied, of the U.S. Department of Homeland Security.

Sandia National Laboratories is a multi-program laboratory managed and operated by Sandia Corporation, a wholly owned subsidiary of Lockheed Martin Corporation, for the U.S. Department of Energy’s National Nuclear Security Administration under Contract DE-AC04-94AL85000. SAND Number 2015-1190 J.

\section*{References}

\bibliography{mybibfile}

\begin{thebibliography}{10}
\expandafter\ifx\csname url\endcsname\relax
  \def\url#1{\texttt{#1}}\fi
\expandafter\ifx\csname urlprefix\endcsname\relax\def\urlprefix{URL }\fi
\expandafter\ifx\csname href\endcsname\relax
  \def\href#1#2{#2} \def\path#1{#1}\fi

\bibitem{Kno2000}
G.~F. Knoll, Radiation Detection and Measurement, 3rd Edition, John Wiley \&
  Sons, Inc., 2000.

\bibitem{Kornilov2003}
N.~Kornilov, V.~Khriatchkov, M.~Dunaev, A.~Kagalenko, N.~Semenova, V.~Demenkov,
  A.~Plompen, Neutron spectroscopy with fast waveform digitizer, Nuclear
  Instruments and Methods in Physics Research Section A 497~(2{\-}3) (2003) 467
  -- 478.
\newblock \href
  {http://dx.doi.org/http://dx.doi.org/10.1016/S0168-9002(02)01790-4}
  {\path{doi:http://dx.doi.org/10.1016/S0168-9002(02)01790-4}}.

\bibitem{Shea2011}
D.~A. Shea, D.~Morgan, The helium-3 shortage: Supply, demand, and options for
  congress, Tech. Rep. R41419, Congressional Research Service (2010).

\bibitem{Adams1978}
J.~Adams, G.~White, A versatile pulse shape discriminator for charged particle
  separation and its application to fast neutron time-of-flight spectroscopy,
  Nuclear Instruments and Methods 156~(3) (1978) 459 -- 476.
\newblock \href
  {http://dx.doi.org/http://dx.doi.org/10.1016/0029-554X(78)90746-2}
  {\path{doi:http://dx.doi.org/10.1016/0029-554X(78)90746-2}}.

\bibitem{Pawelczak2013}
I.~A. Pawe{\l}czak, S.~A. Ouedraogo, A.~M. Glenn, R.~E. Wurtz, L.~F. Nakae,
  Studies of neutron{\-}{$\gamma$} pulse shape discrimination in ej-309 liquid
  scintillator using charge integration method, Nuclear Instruments and Methods
  in Physics Research Section A 711 (2013) 21 -- 26.
\newblock \href
  {http://dx.doi.org/http://dx.doi.org/10.1016/j.nima.2013.01.028}
  {\path{doi:http://dx.doi.org/10.1016/j.nima.2013.01.028}}.

\bibitem{Liao2014}
C.~Liao, H.~Yang, n/{$\gamma$} pulse shape discrimination comparison of ej301
  and ej339a liquid scintillation detectors, Annals of Nuclear Energy 69 (2014)
  57 -- 61.
\newblock \href
  {http://dx.doi.org/http://dx.doi.org/10.1016/j.anucene.2014.01.039}
  {\path{doi:http://dx.doi.org/10.1016/j.anucene.2014.01.039}}.

\bibitem{Pozzi2013}
S.~A. Pozzi, M.~M. Bourne, S.~D. Clarke, Pulse shape discrimination in the
  plastic scintillator ej-299-33, Nuclear Instruments and Methods in Physics
  Research Section A 723 (2013) 19 -- 23.
\newblock \href
  {http://dx.doi.org/http://dx.doi.org/10.1016/j.nima.2013.04.085}
  {\path{doi:http://dx.doi.org/10.1016/j.nima.2013.04.085}}.

\bibitem{Gamage2011}
K.~A.~A. Gamage, M.~J. Joyce, N.~P. Hawkes, A comparison of four different
  digital algorithms for pulse-shape discrimination in fast scintillators,
  Nuclear Instruments and Methods in Physics Research Section A 642~(1) (2011)
  78 -- 83.
\newblock \href
  {http://dx.doi.org/http://dx.doi.org/10.1016/j.nima.2011.03.065}
  {\path{doi:http://dx.doi.org/10.1016/j.nima.2011.03.065}}.

\bibitem{Kaplan2013}
A.~C. Kaplan, M.~Flaska, A.~Enqvist, J.~L. Dolan, S.~A. Pozzi, Ej-309 pulse
  shape discrimination performance with a high gamma-ray-to-neutron ratio and
  low threshold, Nuclear Instruments and Methods in Physics Research Section A
  729 (2013) 463 -- 468.
\newblock \href
  {http://dx.doi.org/http://dx.doi.org/10.1016/j.nima.2013.07.081}
  {\path{doi:http://dx.doi.org/10.1016/j.nima.2013.07.081}}.

\bibitem{Polack2013}
J.~K. Polack, M.~Flaska, A.~Enqvist, S.~A. Pozzi, A computer-aided, visual,
  charge-integration, pulse-shape-discrimination method for organic
  scintillators, in: Proceesdings of the Institute of Nuclear Materials
  Management 54{$^{th}$} Annual Meeting, Palm Desert, California, 2013.

\bibitem{Kauzes2009}
R.~T. Kauzes, J.~R. Ely, A.~T. Lintereur, D.~L. Stephens, Neutron detector
  insensitivity criteria, Tech. Rep. PNNL-18903, Pacific Northwest National
  Laboratory (2009).

\bibitem{Vega2001}
H.~R. Vega-Carrillo, E.~Manzanares-Acuña, A.~M. Becerra-Ferreiro,
  A.~Carrillo-Nuñez, Neutron and gamma-ray spectra of 239pube and 241ambe,
  Applied Radiation and Isotopes 57~(2) (2002) 167 -- 170.
\newblock \href
  {http://dx.doi.org/http://dx.doi.org/10.1016/S0969-8043(02)00083-0}
  {\path{doi:http://dx.doi.org/10.1016/S0969-8043(02)00083-0}}.

\bibitem{Lawrence2014}
C.~C. Lawrence, M.~Febbraro, T.~N. Massey, M.~Flaska, F.~Becchetti, S.~A.
  Pozzi, Neutron response characterization for an ej299-33 plastic
  scintillation detector, Nuclear Instruments and Methods in Physics Research
  Section A: Accelerators, Spectrometers, Detectors and Associated Equipment
  759~(0) (2014) 16 -- 22.
\newblock \href
  {http://dx.doi.org/http://dx.doi.org/10.1016/j.nima.2014.04.062}
  {\path{doi:http://dx.doi.org/10.1016/j.nima.2014.04.062}}.

\bibitem{PeakFit}
T.~O'Haver,
  \href{http://terpconnect.umd.edu/~toh/spectrum/InteractivePeakFitter.htm}{Peak
  fitter version 5.4} (2014).
\newline\urlprefix\url{http://terpconnect.umd.edu/~toh/spectrum/InteractivePeakFitter.htm}

\bibitem{Uchida2014}
Y.~Uchida, E.~Takada, A.~Fujisaki, M.~Isobe, K.~Ogawa, K.~Shinohara, H.~Tomita,
  J.~Kawarabayashi, T.~Iguchi, A study on fast digital discrimination of
  neutron and gamma-ray for improvement neutron emission profile measurement,
  The Review of Scientific Instruments 85~(11) (2014) 11E118.
\newblock \href {http://dx.doi.org/10.1063/1.4891711}
  {\path{doi:10.1063/1.4891711}}.

\end{thebibliography}

\end{document}